\documentclass[preprint,showpacs,preprintnumbers,nofootinbib,amsmath,amssymb,aps,prc,longbibliography]{revtex4-2}
\usepackage{graphicx}
\usepackage{bm}
\usepackage{amsmath}
\usepackage{amssymb,amsthm}
\usepackage[intlimits]{mathtools}
\usepackage{multirow}
\usepackage{float}
\usepackage[colorlinks=true,linkcolor=blue,citecolor=blue,urlcolor=blue,bookmarks=false,hypertexnames=true]{hyperref} 
\begin{document}
	\title{Possible existence of pygmy dipole resonance built on excited states in a neutron-rich $^{80}$Ge nucleus}
	
\author{L. Tan Phuc$^{1,2}$}
\email{letanphuc2@duytan.edu.vn}
\author{N. Dinh Dang$^{3}$}
\author{R. Li$^{4}$}
\author{N. Quang Hung$^{1,2}$}
\email{nguyenquanghung5@duytan.edu.vn}

 \affiliation{
 1) Institute of Fundamental and Applied Sciences, Duy Tan University, Ho Chi Minh City 70000, Vietnam \\
2) Faculty of Natural Sciences, Duy Tan University, Da Nang City 55000, Vietnam\\
3) Nuclear many-body theory laboratory, RIKEN Nishina Center for Accelerator-Based Science, 2-1 Hirosawa, Wako City, 351-0198 Saitama, Japan\\
4) KU Leuven, Instituut voor Kern- en Stralingsfysica, Celestijnenlaan 200D, B-3001 Leuven, Belgium
}
 
	\date{\today}
	
	\begin{abstract}
The pygmy dipole resonance (PDR) at the low-lying tail of the giant dipole resonance (GDR) is an interesting research subject as it carries important information about the nuclear surface with mixed isoscalar and isovector vibrations in $N > Z$ systems. The present paper investigates the possible existence of the PDR built on excited states in a neutron-rich $^{80}$Ge nucleus using the phonon damping model (PDM) with and without pairing correlation at zero and finite temperatures. The results obtained within the PDM with exact pairing (EP+PDM) reveal the appearance of two enhanced $\gamma$-transitions $E_\gamma = 7.25$ and 7.35 MeV at $T=0.6$ MeV, which carry the PDR nature and hence implying the possible existence of PDR built on excited states in this nucleus. These two $\gamma$-energies nicely match with a recent experimental observation, thus indicating the crucial role of the inclusion of exact pairing solution in the precise description of low-lying $\gamma$-transitions. The partition function-based analysis shows that this predicted PDR at $T=$ 0.6MeV is mainly contributed by the first $2^+$ excited state of the $^{80}$Ge nucleus. The isospin mixing at the nuclear surface is also observed in the investigated PDR at $T=0.6-0.7$ MeV. The primary mechanism underlying the emergence of the hot PDR is found due to the coupling of GDR phonon to non-collective particle-particle and hole-hole configurations at finite temperatures within the PDM framework, along with the shift in particle-hole excitation energies due to thermal pairing correlations
	\end{abstract}
	
	\maketitle
	\section{Introduction}
	\label{Intro}
   
The low-lying excited states of the nucleus offer crucial insights into the nuclear structure and reactions, particularly through electromagnetic transitions. These transitions provide essential information on the nuclear resonances, notably the dipole resonances such as the giant dipole resonance (GDR) and/or pygmy dipole resonance (PDR), depending on the energy region of interest. While the GDR has long been known and widely studied \cite{Rowe, Ring}, both theoretical \cite{Lanza} and experimental \cite{Savran} research on the PDR were initiated later and have attracted significant attention in recent years due to their impact on the nucleosynthesis processes in astrophysical environments \cite{Burbidge, Goriely, Goriely2, Qian, Tsoneva, Tonchev}.

The PDR is characterized by the oscillation of excess neutrons against the isospin symmetric core \cite{Bracco}. The PDR states manifested in the vicinity just below the neutron separation energy S$_n$ of neutron-rich nuclei and are known to situate at the low-energy tail of the GDR \cite{Paar, Lanza, Savran, Tsoneva}. Examinations of the oscillation phase via nucleon transition densities indicate that the characteristics of a PDR are the in-phase oscillation of transition densities in the nuclear interior and out-of-phase oscillation at the nuclear surface, where only the neutron transition density contributes. This results in a mixing between the isovector and isoscalar modes, thereby facilitating the experimental investigations of the PDR using both isoscalar and isovector probes \cite{Lanza, Savran, Bracco}. On the other hand, given its sensitivity to the nuclear surface, the PDR is frequently linked to the symmetry-energy and neutron-skin thickness studies \cite{Goriely, Klimkiewicz, Carbone, Reinhard, Piekarewicz, Piekarewicz1, Roca, Papa}. Moreover, the study of the PDR becomes more interesting when considering its contribution to astrophysical processes. For instance, the low-lying resonances like the PDR exert a significant influence on the radiative neutron-capture rate \cite{Paar,Goriely} and play a crucial role in \textit{r}-process calculations, such as the \textit{r}-abundance distribution \cite{Burbidge, Goriely, Goriely2, Qian, Tsoneva}. These resonances also contribute to the \textit{s}-process \cite{Tonchev,Pogliano}, where the medium and heavy elements are formed. This underscores the significance of PDR contributions to the astrophysical observables such as Maxwellian-averaged cross section and reaction rate \cite{Pogliano}, thereby illuminating the origin and the abundance of elements in our universe.

In the context of not having a systematic experimental view on PDR due to technical challenges and device sensitivities \cite{Savran}, there has been a notable push towards the development of theoretical models to describe the PDR. Those theoretical models are generally constructed from both macroscopic and microscopic perspectives. Macroscopic models, such as the hydrodynamical approach \cite{SJ, Mohan, Isacker} or elastodynamic excitation mechanism \cite{Bastrukov}, treat the nucleus as either a liquid droplet or a core-layer structure. In the former, the PDR arises from the oscillations between the neutron fluid at the surface and the symmetric core through compressional modes, whereas in the latter, the PDR is characterized by a nuclear core-layer undergoing elastic-shear vibrations. On the other hand, microscopic models such as Hartree-Fock plus random-phase approximation (HFRPA) \cite{Ring}, renormalized RPA (RRPA) \cite{Hara, Rowe1, Catara, Phuc}, quasiparticle RPA (QRPA) \cite{Ring}, relativistic RPA and QRPA \cite{Wa, Nik, Paar1}, second RPA (SRPA) and subtracted SRPA (SSRPA) \cite{Yan, Tse, Grasso}, multi-phonons models \cite{Bohr, Hage}, quasiparticle phonon model (QPM) \cite{Soloviev, Bertulani}, quasiparticle time blocking approximation (QTBA) \cite{Tse1}, relativistic QTBA (RQTBA) \cite{Lit1, Lit2, Lit3}, equation of motion phonon method (EMPM) \cite{Knapp, Gre}, and phonon damping model (PDM) \cite{Dang, Dang1}, are derived based mainly on the particle-hole ($ph$) residual interactions via the nuclear mean field. Comprehensive reviews of these theoretical models can be found in Refs. \cite{Paar, Lanza, Bracco,Chak} and references therein.

Most of the above-mentioned models only describe the PDR at zero temperature, implying that this resonance is built on the nuclear ground state. This description is based on the conventional point of view that the PDR should come from the low-lying $1^-$ excited states de-exciting to the ground state. It is naturally expected that there should also be the PDR built on excited states or the PDR at finite temperatures (hot PDR). At finite temperature, the nuclear system is known to be in a thermodynamical ensemble of many excited states at thermal equilibrium \cite{HungRep}. Thus, the study of hot PDR is equivalent to the study of the PDR built on excited states. From the theoretical perspective, such a study is possible since the PDR is an inseparable part of the GDR, which is influenced by temperature. Many theoretical and experimental studies have proved this influence via the damping of hot GDR \cite{Dang, Gaar, Snover, Faou, Bau, Rama, Harakeh, Bor, San, Chak}.  It can also be indirectly studied via the enhancement of the radiative $\gamma$-rays strength function (RSF) in the PDR energy region, which has been observed in recent years \cite{Voinov, Gutt, Toft, Erik}. Theoretical RSF models can describe such enhancements only if the temperature effect is included \cite{Voinov, Gutt, KMF, HungPRL, Phuc1, Goriely3}. 

Recent experiments have reported the decay patterns of the PDR in several nuclei, in which a significant portion ($\sim$ 25\%) of the decay to low-lying excited states were observed \cite{Angell, Scheck, Isaak, Loher}. Another experiment has observed the extra $\gamma$-ray yields from the tails of the hot GDR in neutron-rich $^{60,62}$Ni nuclei \cite{Wie}. Those extra yields may be attributed to the hot PDR. In particular, a very recent experiment based on the $\beta$-decay of a neutron-rich $^{80}$Ge nucleus has reported the first evidence of the existence of PDR built on excited states with non $1^-$ spin-parity \cite{Ren}. This experiment has observed two $\gamma$-decays of 7181(53$_{stat}$ + 18$_{sys}$) and 7337(53$_{stat}$ + 19$_{sys}$) keV associated, respectively, to the direct transitions from 7840 and 7796 keV excited states to the low-lying $2_1^+$ states of $^{80}$Ge, thus implying the first signature of a PDR built on non $1^-$ excited states. Those experimental studies motivate our theoretical investigation on the hot PDR in neutron-rich nuclei.

The goal of the present paper is to investigate the possible existence of PDR built on excited states in the above $^{80}$Ge nucleus by using the PDM plus exact pairing at finite temperature (EP+PDM). The PDM has been proposed and successfully employed to describe the evolution of the GDR width as a function of temperature in various hot nuclei \cite{Dang,Dang1,Chak}. When coupling with the exact pairing (EP) solution, the PDM is able to describe reasonably well the RSFs of several nuclei \cite{HungPRL,Phuc1,PhucPLB}. In addition, the observation of decay modes of the PDR mentioned above suggests the important contribution of higher-order excitations, such as $2p2h$ and $3p3h$, to the PDR formation. In this context, the PDM and EP+PDM, which take into account the couplings of the GDR phonon to the $ph$, $pp$, and $hh$ configurations, are equivalent to the couplings to $2p2h$ configurations, and thus suitable to describe the hot PDR. The obtained results may provide more insights into the structure of hot PDR and inspire further experimental search. The paper is organized as follows. After the Introduction in Sec. \ref{Intro}, the PDM and EP+PDM formalisms are briefly presented in Sec. \ref{Forma}. The results obtained are discussed in Sec. \ref{Resu}.  Conclusions are drawn in the last section.
	
\section{Formalism}
\label{Forma}
In this work, two versions, conventional PDM (without pairing) \cite{Dang} and EP+PDM (with exact pairing) \cite{DangPRC2012}, are employed to calculate the electric dipole ($E1$) strength function/distribution. The PDM and EP+PDM formalisms have been presented in detail in several publications, e.g. Refs. \cite{Dang,Chak,HungPRL,Phuc1,DangPRC2012,Dang2}, so we only summarize the main idea of these models. 

The PDM is derived based on a Hamiltonian containing three terms, namely the independent single-particle (quasiparticle) field with single-particle (quasiparticle) energies $\epsilon_k (E_k)$, the phonon field with phonon energies $\omega_q$, and the coupling between these two fields, which causes the damping of the GDR \cite{Dang,DangPRC2012}. Within the PDM, the $E1$ strength distribution at a given $\gamma$-energy $E_\gamma$ and temperature $T$ is calculated as 
\begin{equation}
\label{SE1}
S_{E1}(E_{\gamma},T)=\frac{1}{\pi}\frac{\gamma_q(E_{\gamma},T)}{(E_{\gamma}-E_{\rm GDR})^2+\gamma^2_q(E_{\gamma},T)}~,
\end{equation}
where $E_{GDR}$ is the temperature-independent GDR energy and $\gamma_q(E_\gamma,T)$ is the phonon damping, which is related to the GDR full width at half maximum $\Gamma_{\rm GDR}(T)$ by the following relation 
\begin{equation}
\gamma_q(E_\gamma=E_{\rm GDR},T) = \frac{1}{2}\Gamma_{\rm GDR}(T) ~.
\end{equation}
At $T=0$, the phonon damping $\gamma_q(E_\gamma)$ is microscopically calculated within the PDM by coupling the GDR phonon to the collective $ph$ states. It corresponds to the quantal width of the GDR, that is the width of the GDR built on the ground state. At $T\neq 0$, the PDM additionally includes the coupling of the GDR phonon to the non-collective $pp$ and $hh$ states, which causes the thermal width of the GDR. As the temperature increases, the quantal width slightly decreases, whereas the thermal width increases and saturates at a sufficiently high $T$. This is how the PDM explains the evolution of the GDR width as a function of temperature (See \cite{Dang} for details). The explicit expression of $\gamma_q(E_\gamma,T)$ is given as the sum of these quantal and thermal half-widths as follows
\begin{eqnarray}
\hspace{-5mm} \label{gammaPDM}
\gamma_q(E_\gamma,T) &=& \pi F_{ph}^2\sum_{ph}[u_{ph}^{(+)}]^2(1-n_p-n_h)\delta(E_\gamma-E_p-E_h) ~\nonumber \\
&+&\pi F_{ss'}^2\sum_{s>s'}[v_{ss'}^{(-)}]^2(n_s-n_{s'})\delta(E_\gamma-E_s+E_{s'})~,
\end{eqnarray}
where $ss'$ stands for $pp'$ and $hh'$, whereas $u_{ph}^{(+)}=u_p v_h + v_p u_h$ and $ v_{ss'}^{(-)}=u_s u_{s'} - v_s v_{s'} $ are the combination of the Bogoliubov's coefficients $u_k$ and $v_k$ ($k=p,h$). The $\delta$ function in Eq. \eqref{gammaPDM} is given the form of $\delta(x)=\dfrac{\varepsilon}{\pi(x^2+\varepsilon^2)}$ with $\varepsilon$ being the smoothing parameter. The quasiparticle occupation numbers $n_k$ are given in terms of the Fermi-Dirac distribution 
\begin{equation}
n_k=\frac{1}{e^{E_k/T}+1} ~,
\end{equation}
where $E_k=\sqrt{[\epsilon_k-\lambda(T)]^2+\Delta(T)^2}$ are the quasiparticle energies with $\lambda(T)$ and $\Delta(T)$ being the temperature-dependent chemical potential and pairing gap, respectively. Within the PDM (without pairing correlation), the pairing gap $\Delta(T)$ is zero, thus the quasiparticle energies $E_k$ become $E_k=|\epsilon_k-\lambda(T)|$ and $\lambda(T)$ can be obtained from the independent-particle model (IPM) at finite temperature. When pairing correlation is taken into account, all the coefficients $u_k, v_k, E_k$, and $n_k$ are calculated by using the EP and the model becomes the EP+PDM. In the latter, the EP is treated for a selected number of levels around the Fermi surface (truncated levels), where pairing correlation is known to be most effective. Beyond the truncated levels, where pairing correlation has negligible effect, the finite-temperature independent-particle model is used \cite{HungPRL,DangPRC2012,Dang2}. The coupling matrix elements $F_{ph}, F_{pp'}$, and $F_{hh'}$ are set to be constant within the PDM or EP+PDM, namely $F_{ph}=F_1$ and $F_{pp'}=F_{hh'}=F_2$. Those $F_1$ and $F_2$ are considered as the model parameters. The value of $F_1$ is adjusted to reproduce the experimental GDR width at $T=0$ or that determined from the global parameter \cite{ripl2} if no experimental width is found. The $F_2$ value is also selected at $T=0$ so that the calculated GDR energy is insignificantly changed with $T$ \cite{Dang1,HungPRL,PhucPLB}.

To look deeper into the GDR and PDR structures, we calculate the transition densities in the corresponding GDR and PDR regions. The transition density within the PDM and EP+PDM has the form
\begin{equation}
\label{dens}
\delta\rho_{\rm trans}(r)=\frac{1}{\sqrt{4\pi}}\frac{1}{\sqrt{2J+1}}\sum_{kk'}(-1)^{l_k+l_{k'}}\frac{\gamma_{kk'}(E_\gamma,T)}{F_{kk'}}\frac{\varphi_k(r)\varphi_{k'}(r)}{r^2}~,
\end{equation}
which is similar to that used in the Skyrme HFRPA in Ref. \cite{Colo}. In Eq. \eqref{dens}, $\varphi_{k(k')}(r)$ are the radial single-particle wave functions of the $k(k')$-th single-particle levels. The term $\gamma_{kk'}(E_\gamma,T)/F_{kk'}$, which plays the same role as the RPA matrix element, stands for the contribution of the $kk’$ transition strengths to the total transition density at the energy $E_\gamma$ and temperature $T$. This quantity is obtained within the PDM or EP+PDM using Eq. \eqref{gammaPDM}. 

\section{Results and Discussion}
\label{Resu}

\begin{figure}[!h]
 \includegraphics[width=1\textwidth]{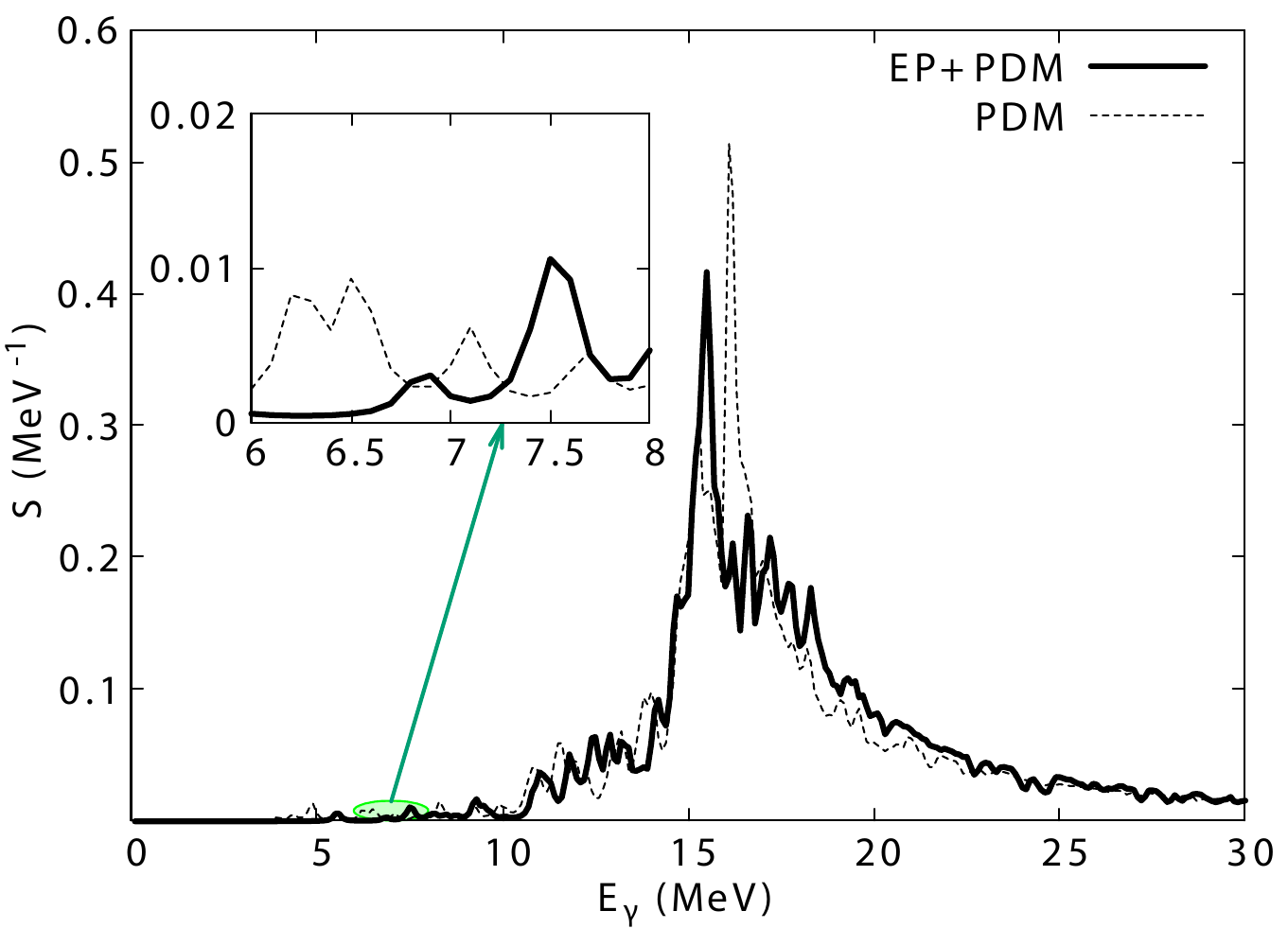}
 \caption{\label{fig1} The $E$1 strength functions of $^{80}$Ge obtained within the PDM and EP+PDM at $T = 0$ by using the value of smoothing parameter $\varepsilon=0.1$ MeV.}
 \end{figure}

The PDM and EP+PDM are used to calculate the $E$1 strength distribution of $^{80}$Ge nucleus. The neutron and proton single-particle spectra are calculated by using the Skyrme Hartree-Fock mean field \cite{Colo}. The SLy5 interaction is employed for illustration. The obtained results do not significantly change with the different version of the interactions. The obtained single-particle spectra and wave functions are then used to calculate the $E$1 strength functions \eqref{SE1} and transition densities \eqref{dens}. Since $^{80}$Ge is slightly deformed (the quadrupole deformation parameter $\beta_2 = 0.182$ \cite{NNDC}), its GDR line shape is split into two components, whose energies and widths can be retrieved from the RIPL-3 nuclear database \cite{RIPL3}, namely at $E_1^I \sim 15.9$ MeV, $E_1^{II} \sim 18.2$ MeV, $\Gamma_1^I \sim 4.11$ MeV, and $\Gamma_1^{II} \sim 6.40$ MeV. Once the GDR is fixed, the low-lying $E$1 transitions can be observed from the GDR tail below S$_n$ as shown in Fig. \ref{fig1} for $T = 0$ case. The energy region used to investigate the PDR is from $6-8$ MeV as suggested by the latest experiment in Ref. \cite{Ren}. Although deformation influences the GDR, causing it to split into a double-bump structure, this splitting in the PDR region is not clearly observed in our study. This is consistent with the results reported in Ref. \cite{Lanza}

In Fig. \ref{fig1}, one can clearly observe the appearance of resonances at $E_\gamma \sim 6 - 8$ MeV within both PDM and EP+PDM. Within the PDM, there appears a strong resonance, whose peak position is located at $E_\gamma \sim 6.5$ MeV. This resonance is shifted to $E_\gamma \sim 7.5$ MeV within the EP+PDM (see the inset of Fig. \ref{fig1}). The wide energy distribution of this resonance indicates the possibility that the resonance is formed by two or more $E$1 transitions. In addition, the shift of the resonance to the higher energy position as that obtained within the EP+PDM is attributed to the effect of exact pairing, which minimizes the system's energy. Consequently, the resonances at lower energies require more energy to occur. Here, we recall that the PDM and EP+PDM calculations in Fig. \ref{fig1} are performed based on the conventional GDR at $T = 0$, namely the $E$1 transitions from the $1^-$ excited states to the ground state. Thus, the resonances observed from the GDR tail might have the same nature, i.e., resonances built on the ground state.

 \begin{figure}[!h]
 \includegraphics[width=1\textwidth]{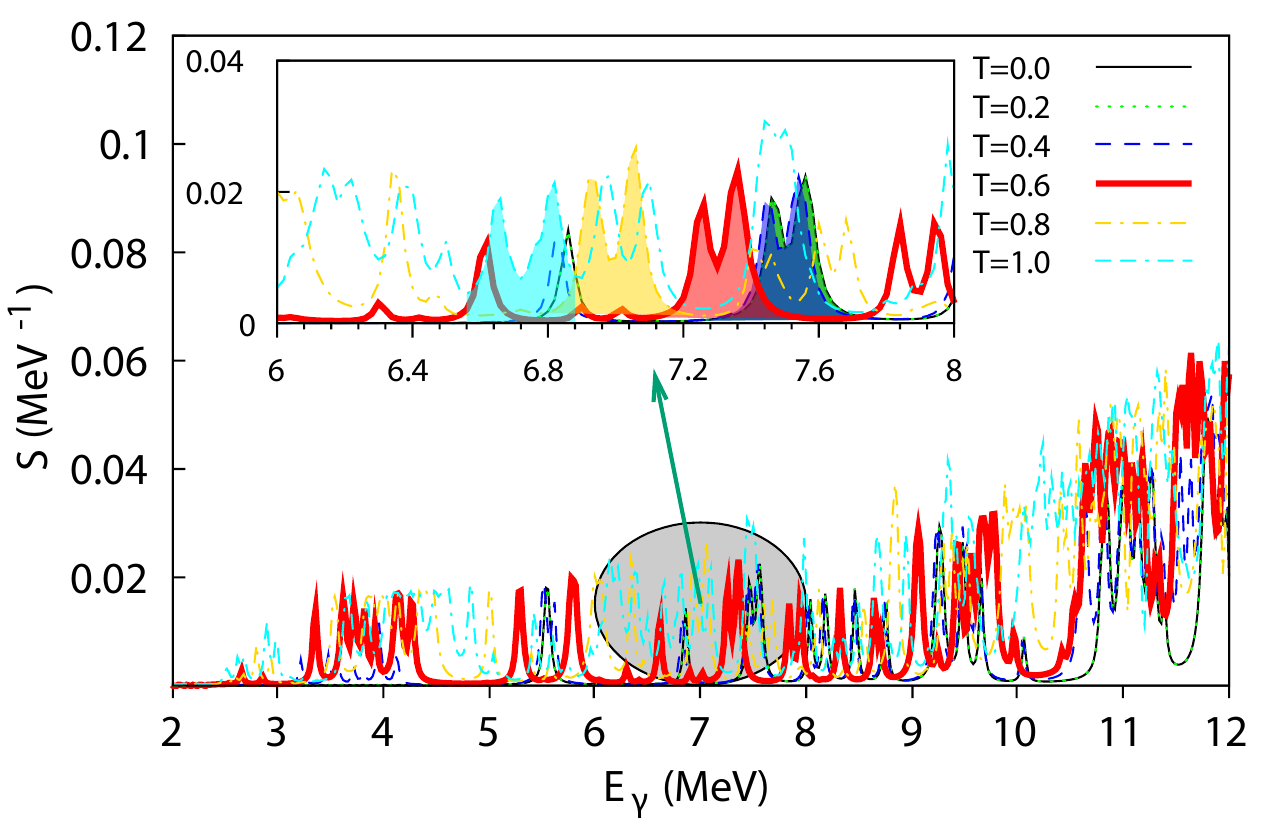}
 \caption{\label{fig2} The $E$1 strength functions of $^{80}$Ge obtained within the EP+PDM at $T = 0 - 1$ MeV by using the value of smoothing parameter $\varepsilon=20$ keV}
 \end{figure}

To study the PDR built on the excited states, we perform the EP+PDM at $T \neq 0$. Fig. \ref{fig2} shows the $E1$ strength functions obtained within the EP+PDM in the region of GDR's low-lying tail at $T = 0 - 1$ MeV. Here, one can see the presence of the resonances at $E_\gamma \sim 6 - 8$ MeV at all considered temperatures. However, with increasing $T$, the peak position of the largest resonance, fully filled by colors in Fig. \ref{fig2}, is shifted towards lower energy due to the thermal effect, which reduces the pairing correlation. For example, at $T = 0.6$ MeV, the peak position of the largest resonance is shifted to $E_\gamma \sim 7.3$ MeV, which is about 0.2 MeV smaller than that obtained from the largest resonance built on the ground state in Fig. \ref{fig1}. Hereafter, the energy region $6-8$ MeV will be carefully analyzed as this is where the experimental observation of the PDR built on excited states was reported \cite{Ren}.

 \begin{figure}[!h]
 \includegraphics[width=1\textwidth]{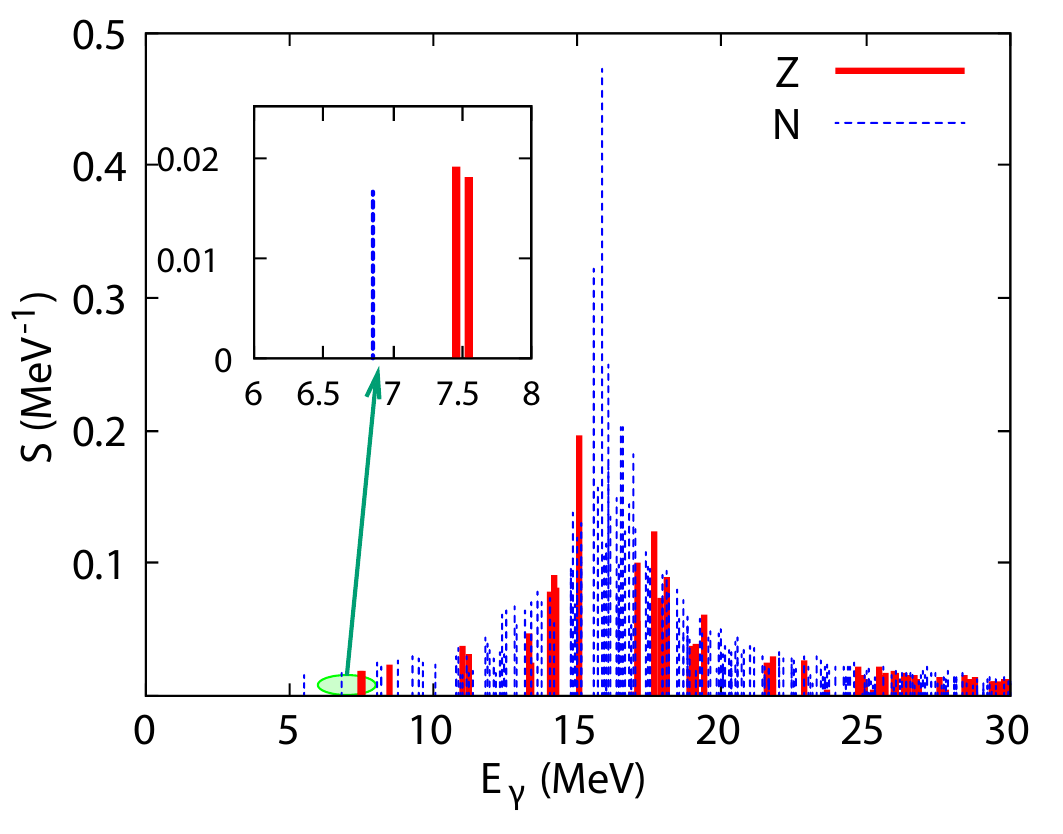}
 \caption{\label{fig3} The proton (Z) and neutron (N) $E$1 strength functions of $^{80}$Ge obtained within the EP+PDM at $T = 0$ MeV by using a very small value of smoothing parameter $\varepsilon=10$ keV.}
 \end{figure}

\begin{figure}[!h]
 \includegraphics[width=1\textwidth]{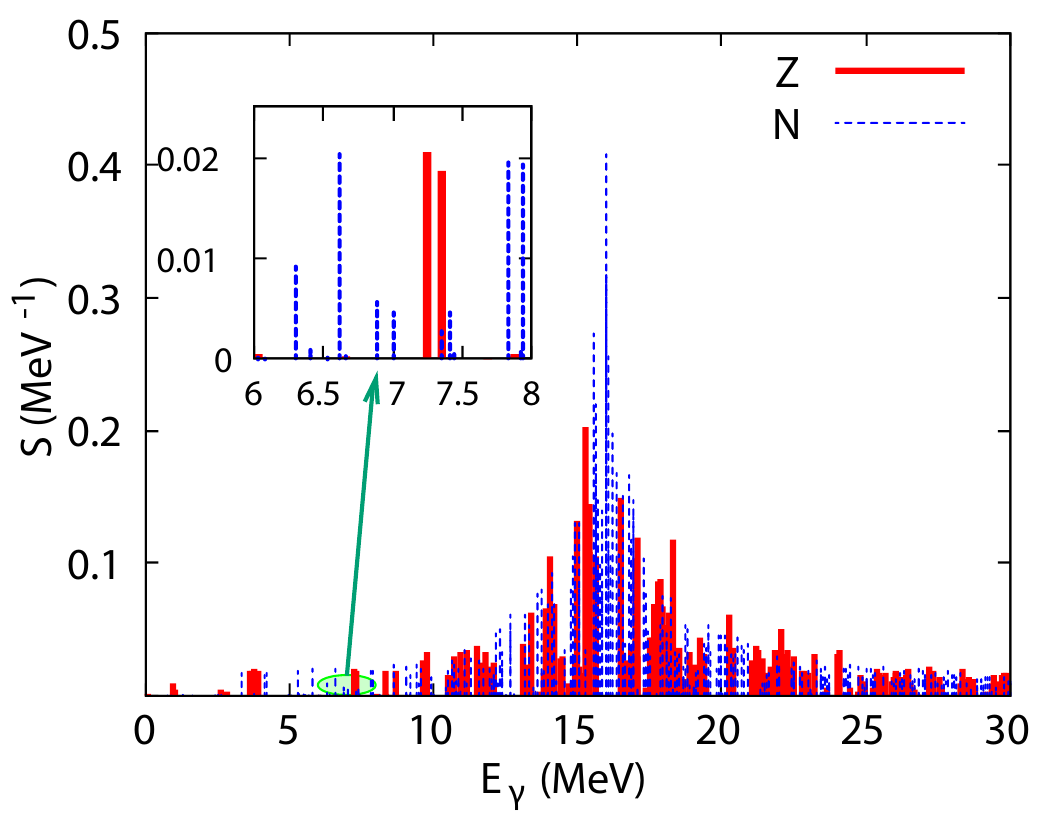}
 \caption{\label{fig4} Same as Fig. \ref{fig3} but at $T=0.6$ MeV}
 \end{figure}

Within the PDM or EP+PDM, the proton and neutron mean fields are separately treated and the total strength function is simply the sum of the strength functions for neutron and proton. Thus, it allows to examine the contribution of proton and neutron strengths to the total one in each resonance. Figs. \ref{fig3} and \ref{fig4} show the proton and neutron $E1$ strength functions obtained within the EP+PDM at $T = 0$ and 0.6 MeV using a very small value of smoothing parameter $\varepsilon$. By separately analyzing the proton and neutron $E$1 strengths that contribute to the total GDR, one can investigate the origin of the resonances in the PDR region. At $T = 0$ MeV, it is evident that in the region of $E_\gamma = 7 - 7.6$ MeV, only proton transitions contribute, yielding two peaks at $E_\gamma \sim 7.45$ and 7.55 MeV (see the inset of Fig. \ref{fig3}). However, at $T = 0.6$ MeV, the neutron transitions emerge alongside the proton ones in this energy region, accompanied by a shift of two proton-dominated resonance peaks towards lower energies $E_\gamma \sim 7.25$ and 7.35 MeV (see the inset of Fig. \ref{fig4}). This result implies that at $T = 0.6$ MeV, there should be at least two enhanced $E$1 transitions at $E_\gamma \sim 7.25$ and 7.35 MeV to some low-lying excited states, not the ground state. Notably, this result is in agreement with the experimental results in Ref. \cite{Ren}, in which two $\gamma$-transitions of 7181(53$_{stat}$+18$_{sys}$) and 7337(53$_{stat}$+19$_{sys}$) keV associated to the PDR built on exited $2_1^+$ states were reported.

 \begin{figure}[!h]
 \includegraphics[width=0.5\textwidth]{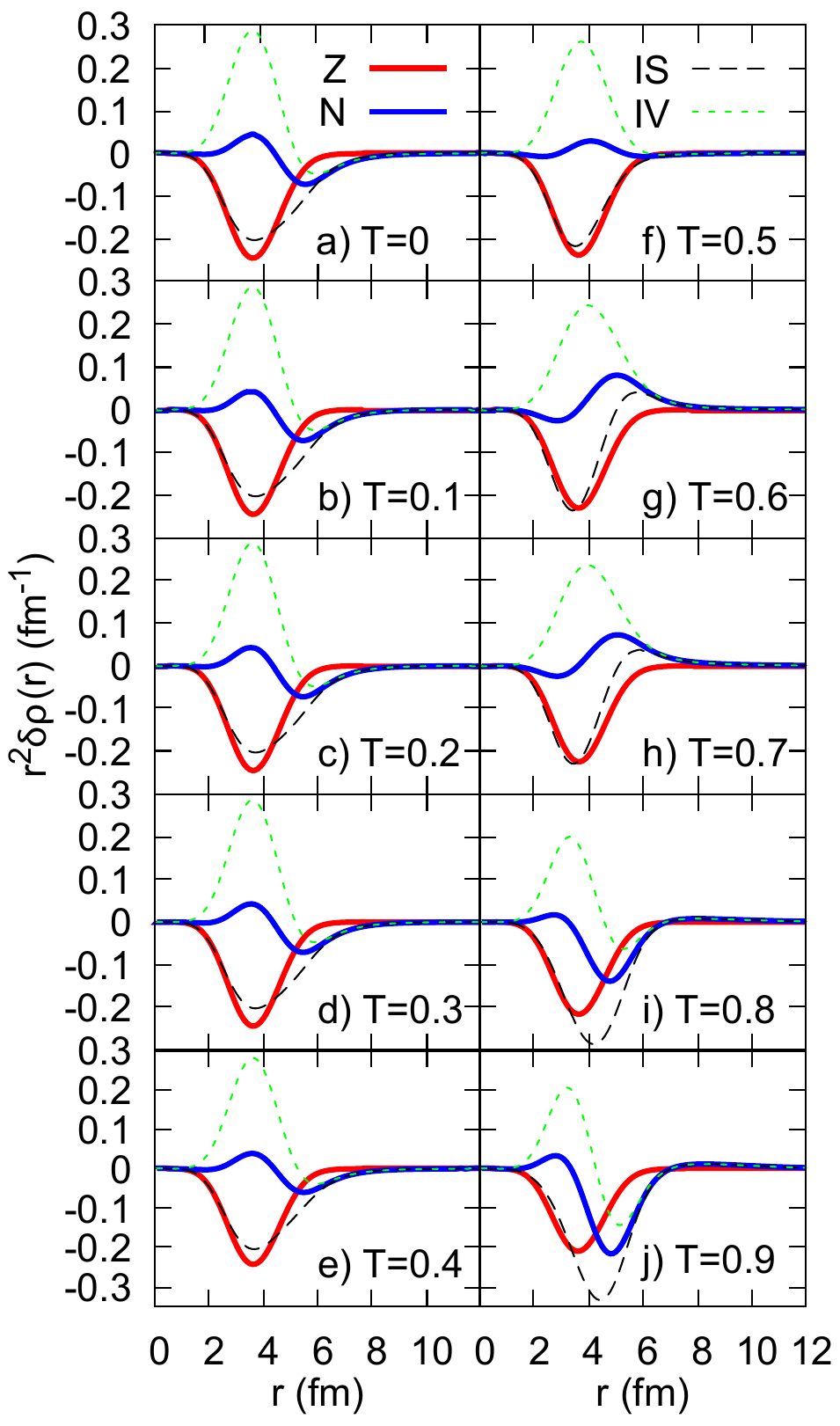}
 \caption{\label{fig5} The neutron and proton transition densities along with their IS and IV modes in the PDR region ($6-8$ MeV) obtained within the EP+PDM for $^{80}$Ge at finite temperatures.}
 \end{figure}
 
 \begin{figure}[!h]
 \includegraphics[width=1\textwidth]{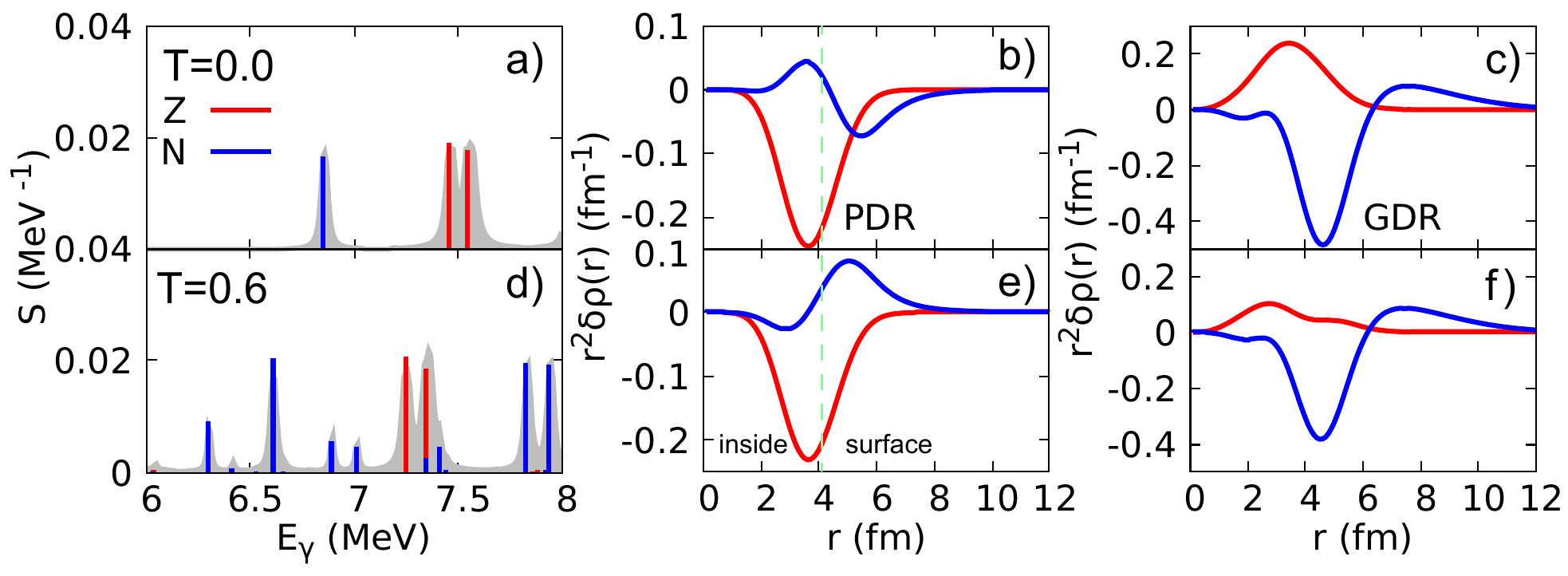}
 \caption{\label{fig6} The neutron and proton strength functions in the PDR region together with the neutron and proton transition densities summed over the PDR and GDR regions obtained within the EP+PDM for $^{80}$Ge at $T = 0$ and 0.6 MeV.}
 \end{figure}

To further investigate the nature of the obtained resonances, we plot in Fig. \ref{fig5} the neutron and proton transition densities in the PDR region ($6-8$ MeV) at finite temperatures. Traditionally, the PDR was considered to be of $ ph $ nature. However, recent studies based on the transition probabilities from the low-lying $1^-$ state to the ground state using the microscopic Skyrme Hartree-Fock RPA \cite{Lanza} have suggested that the PDR is not purely based on $ph$ configurations, and even it cannot be described as a collective resonance. In our EP+PDM model, a similar situation occurs. The GDR phonon, which can be described as collective $ph$ excitations within the microscopic HF+RPA, interacts with single-particle $p$ and $h$ states via its couplings to $ph$, $pp$, and $hh$ configurations at finite temperatures \cite{Dang}, leading to an impurity in the $ph$ nature of the PDR ($ph + pp + hh$). Our present calculations are in good agreement with the findings in Ref. \cite{Lanza}

Fig. \ref{fig5}(a)-(j) show the transition densities of $^{80}$Ge from $T=$0 to 0.9 MeV obtained within the EP+PDM. The signature of the PDR, including in-phase behavior of transition densities in the core and the presence of only neutrons at the nuclear surface, are faintly observed at $T=$0.5 MeV, become pronounced at $T=$0.6 MeV, and then gradually disappear as the temperature become higher than 0.7 MeV. In Fig. \ref{fig5}, we find that in low-temperature region $T = 0-0.5$ MeV, the oscillations in the 6-8 MeV region are more isovector (IV) than isoscalar (IS) because they are simply the tail of the GDR. In the $T = 0.6-0.7$ MeV region, one can see that the IS part increases to be compatible with the IV one. At higher temperatures $0.6 < T < 1.0$ MeV, the IS part increases further and becomes dominant. At $T > 1$ MeV, the IV part becomes dominant. To analyze in detail the typical $T=$0.6 MeV case, we plot in Fig. \ref{fig6} the neutron and proton strength functions in the PDR region together with the neutron and proton transition densities in the PDR and GDR ($\sim 14-16$ MeV) regions obtained within the EP+PDM at $T = 0$ and 0.6 MeV. Fig. \ref{fig6}(a) clearly shows that at $T = 0$, two peaks, whose peak positions at $E_\gamma \sim 7.45$ and 7.55 MeV are found. Those peaks are shifted to $E_\gamma \sim 7.25$ and 7.35 MeV at $T = 0.6$ MeV [Fig. \ref{fig6}(d)]. As for the transition densities, Fig. \ref{fig6}(e) shows a typical PDR nature, namely protons and neutrons are oscillated in-phase in the core ($r < 4$ fm) and out-of-phase near the surface ($4 <  r < 6$ fm). At the surface ($r > 6$ fm), the neutron transition density still survives, while the proton one is almost quenched. This suggests the isospin mixing at the nuclear surface in the investigated PDR, offering the possibility of applying both isovector and isoscalar probes in $^{80}$Ge nucleus, which is in line with previous suggestions \cite{Lanza, Savran, Bracco}. Meanwhile, Figs. \ref{fig6}(c) and (f) reveal a typical GDR nature (proton and neutron oscillations are out-of-phase), in which the neutron oscillations are stronger than the proton ones at $T=0$ as this $^{80}$Ge nucleus is neutron-rich. In the PDR region, the out-of-phase proton-neutron oscillations in the core present at $T = 0$ as seen in Fig. \ref{fig6}(b). It is hard to say if those oscillations have the PDR nature or are simply remnants of the GDR. However, the neutron and proton oscillations are compatible at $T = 0.6$ MeV (Fig. \ref{fig6}(e)). Concerning the variation of neutron transitions at $T =$ 0.6 MeV mainly about the phase change, generally, it arises from the presence of neutron $pp$ (dominant) and $hh$ (minor) excitations within the EP+PDM at finite temperature together with the shift of $ph$ excitations to the lower energy region due to the thermal pairing correlation. This phenomenon occurs only at finite temperature. The phase changes in the PDR region can be seen in Fig. \ref{fig5}. Meanwhile, the weakening of transition densities especially for protons in Fig. \ref{fig6}(f) is simply explained by the thermal damping of the GDR phonon, which is effectively described within the EP+PDM.

To determine which excited states the PDRs are built on, we analyze the EP+PDM thermal equilibrium background, on which the transitions are built at finite temperatures. To do this, the partition function $Z$ of the system is constructed within the EP+PDM. However, to ensure the multiplicity of the system, the partition function must be built on all microstates. This will make the partition function realistic and can correctly describe the thermodynamic quantities of the system such as free energy, heat capacity, and entropy \citep{HungPRL}. Since the EP method is calculated using only a truncated space as mentioned in Sec. \ref{Forma}, $Z_{EP}$ is incomplete. To supplement it, we take into account the contribution of the levels outside the truncated space using the finite-temperature independent-particle model (IPM), namely
\begin{equation}
Z(T)=Z_{EP}(T)+Z_{IPM}(T) .
\end{equation}
The explicit expressions of these functions can be found in Refs.\cite{Hung2009,Alhassid2003}. The contribution of the excited states of $^{80}$Ge to the EP+PDM thermal equilibrium can be evaluated based on the ratio between the partial partition function of excited states and the total partition function as follows
\begin{equation}
C= \frac{(2J+1)e^{-E(J^\pi)/T}}{Z(T)},
\end{equation}
where $E(J^\pi)$ is the excited energy of state $J^\pi$ collected from the experimental level scheme on NNDC database \cite{NNDC}. The results obtained are presented in Table I, which indicates the percentage contribution of the ground state (g.s.) and excited states $J^\pi$ to the EP+PDM thermal equilibrium background in the temperature range $T =$ 0 - 1 MeV. At $T=0$,  all the transitions (100\%) fall to the ground state. As temperature increases, this ratio rapidly decreases, with more transitions distributed to higher excited states. The higher the temperature goes, the larger the contribution from these excited states reaches. In the range of $T =$ 0.6 - 1 MeV, the contribution of the ground state drops continuously below 20\% and becomes difficult to accurately estimate due to the increasing contribution of higher excited states, which are not taken into account in our estimation. Nevertheless, as seen in Table I, at $T = $ 0.6 MeV where the PDR emerges, the contribution of the first $2^+$ excited state is dominant ($\sim 50 \%$). This means that the PDRs are mainly contributed to this $2^+_1$ state, which is consistent with the conclusion in Ref. \cite{Ren}.

\begin{table}
\caption{The contribution $C(\%)$ of excited states of $^{80}$Ge into the EP+PDM thermal equilibrium background at finite temperatures. The unit of excited energies and temperatures are in MeV.}
\begin{tabular}{ccccccc}
\hline \hline
T & $ g.s. $ & $ E(2_1^+)=0.659 $ & $ E(2_2^+)=1.573 $ & $ E(4_1^+)=1.742 $ & $ E(6_1^+)=2.978 $ & $ E(8_1^+)=3.445 $ \\ 
\hline 
0.0 & 100\% & 0\% & 0\% & 0\% & 0\% & 0\% \\ 
0.1 & 99.7\% & 0.3\% & 0\% & 0\% & 0\% & 0\% \\ 
0.2 & 90.6\% & 9.3\% & 0.1\% & 0\% & 0\% & 0\% \\ 
0.3 & 69.7\% & 27.7\% & 1.3\% & 1.3\% & 0\% & 0\% \\ 
0.4 & 43.3\% & 46.2\% & 4.7\% & 5.3\% & 0.4\% & 0.1\% \\ 
0.5 & 22.3\% & 55.3\% & 8.9\% & 11.4\% & 1.4\% & 0.7\% \\ 
0.6 & $<$20\% & 50.4\% & 11.0\% & 14.9\% & 2.7\% & 1.7\% \\ 
0.7 & $<$20\% & 38.3\% & 10.4\% & 14.7\% & 3.6\% & 2.4\% \\ 
0.8 & $<$20\% & 27.1\% & 8.6\% & 12.6\% & 3.9\% & 2.8\% \\ 
0.9 & $<$20\% & 19.1\% & 6.9\% & 10.3\% & 3.8\% & 2.9\% \\ 
1.0 & $<$20\% & 13.7\% & 5.5\% & 8.3\% & 3.5\% & 2.9\% \\ 
\hline 
\end{tabular} 
\centering
\end{table} 

The above analysis suggests that there should be two enhanced $E$1 transitions of $E_\gamma \sim 7.25$ and 7.35 MeV, carrying the PDR nature, to some low-lying excited states in $^{80}$Ge nucleus. Notably, these characteristic transitions are only observed at $T \sim 0.6-0.7 $ MeV. Hence, the present work shows the possibility of the emergence of the PDR at finite temperature in $^{80}$Ge. This finding suggests that the Brink-Axel hypothesis (BAH) \cite{BAH,Axel}, which states that there exists a resonance built on each excited state the same properties as those built on the ground state, is violated in the PDR region. In particular, the PDR appears only within a specific temperature range ($\sim 0.6-0.7$ MeV), driven by the in-phase behavior of proton and neutron transition densities at the nuclear center, thus leading to a violation of the BAH. Other studies have also shown this feature, but at much higher temperatures ($T >$ 3 MeV), where low-lying resonances appear \cite{Lit4,Lit5}. Although there is considerable controversy surrounding the validity of the BAH, recent experimental \cite{Sie} and theoretical \cite{Wasi,Johnson} studies have demonstrated the violation of BAH in the PDR region for lighter nuclei, such as $^{22-27}$Ne, as well as for heavier nuclei like $^{208}$Pb. Our results are consistent with the conclusions of these studies. On the other hand, the partition function-based analyzing results show a high probability that the PDRs are built on the $2^+_1$ state of the $^{80}$Ge nucleus. Our finding is consistent with the experimental results in Ref. \cite{Ren}. However, due to the limitation of the present model, we are unable to explicitly indicate from which excited states come those $\gamma$-transitions. To do so, the model needs further developments, which will be reported in the forthcoming studies.

\section{Conclusion}
The present work investigates the resonances in the low-lying tail of the GDR of a neutron-rich $^{80}$Ge nucleus at zero and finite temperatures. The microscopic phonon-damping model with the exact pairing solution is used for this purpose. The obtained results show the appearance of two $\gamma$-transitions of $E_\gamma \sim 7.25$ and 7.35 MeV at $T=0.6-0.7$ MeV. Those transitions have the PDR nature, thus implying the possible existence of the PDR built on excited states. These PDRs are most likely built on the first $2^+$ excited state of the $^{80}$Ge nucleus. This finding is in good agreement with the latest experimental results in Ref. \cite{Ren}. In addition, the inclusion of exact pairing is crucial to obtain the precise $\gamma$-transitions when compared with the experimental observations. In particular, within the framework of the present model, we found that the coupling of the GDR phonon to collective $ph$ as well as non-collective $pp$ and $hh$ configurations at finite temperature, which causes the increase in the damping of the GDR, along with the shift of particle-hole excitations to the lower energy region due to thermal pairing, are the primary mechanism underlying the emergence of PDR built on excited states. Analysis of transition densities also reveals the isospin mixing character at the nuclear surface in the investigated PDR, which offers the possibility to apply both isovector and isoscalar probes for this $^{80}$Ge nucleus. 

\acknowledgments
The authors (L.T.P and N.Q.H) express their gratitude for all the valuable support from Duy Tan University (DTU). DTU is going to celebrate its 30$^{\rm th}$ anniversary of establishment (Nov. 11, 1994 $-$ Nov. 11, 2024) towards “Integral, Sustainable and Stable Development”. R. Li acknowledges support by KU Leuven postdoctoral fellow scholarship.

\end{document}